\documentclass[amsmath,amssymb,lengthcheck,aps,prl] {revtex4}
\usepackage{graphics}
\usepackage{graphicx}

\begin{document}

\title{Bose-Einstein Condensation in Non-Uniform Rotation}

\author{Saswata Sahu\footnote{SS: saswata.rs2016@physics.iiests.ac.in}, Dwipesh Majumder\footnote{
DM: dwipesh@physics.iiests.ac.in}}
\affiliation{Department of Physics, Indian Institute of Engineering Science and Technology, Shibpur, Howrah, West Bengal, India}

\begin{abstract}
In this work, we would like to present the Bose-Einstein Condensation in such a system where rotation is decreasing radially from the center of the condensate. That non-uniform rotation is defined by a rotating parameter called $\lambda$. The system is defined by a modified Gross-Pitaevskii equation. The result shows very different behavior from uniformly rotating condensate. In an uniformly rotating case there is formation triangular vortex lattice but in our case, we are watching that vortices are formed in circular ring shape above a specific amount rotation defined by $\lambda_c$. Below $\lambda_c$  there is a distortion i.e. there is neither circular ring shape nor triangular symmetry among the vortices. We have studied the energy and chemical potential of the system. We have seen a sharp change in the energy and chemical potential of the systems at the point of $\lambda_c$. In this rich complex phase, we have drawn a phase diagram associated with the phase transition from disordered to circular ring shape pattern.
\end{abstract}

\maketitle

\section{I. Introduction}
The Bose-Einstein Condensation (BEC) becomes one of the intriguing phenomena after the experimental achievement of BEC in low density gas of $^8$$^7$Rb atoms confined in an optically trapped nearly absolute zero temperature\cite{Wieman}  by Wieman and Cornell in 1995. After this breakthrough BEC is also achieved in other alkali atoms like $^2$$^2$Na \cite{sodium}, $^5$$^2$Cr \cite{cromium}, $^7$Li \cite{lith}, $^1$$^3$$^3$Cs \cite{cesium}  etc.
A beautiful Abrikosov vortex lattice (triangular lattice) forms in trapped rotating BEC which have been reported by several experimental groups such as the MIT group \cite{sodium} in 1995, JILA group \cite{35, 57} in 1999, the ENS group \cite{56} in 2000 etc. This vortex lattice has also been established theoretically by several researchers \cite{wbao rot, spectral}.

Vortex lattice has many diverse phases, which have been studied for the last two decades. In two dimension system, one can observe the phase transitions in an interacting system at finite temperature ($T > 0$). One of the remarkable phenomenon is the BKT phase transitions which is the transition from bound vortex-antivortex pair to free vortices named after Berezinskii, Kosterlitz and Thouless \cite{pitsing}. Richard J. Fletcher \textit{et al.} have shown that BKT transition smoothly converges onto BEC \cite{BKTBEC}. It is quite an unconventional phase transition which doesn't break continuous symmetries. Another phenomenon studied by the researchers is Tkachenko oscillations. It is the oscillation of vortex centers in rapidly rotating condensate \cite{Tkachenko}. The curvilinear rows of vortex centers go through the center of the cloud and fit with sine curve very well. At very low temperature the thermal fluctuation is very less but the quantum fluctuation is high. So, in this situation, the microscopic fluctuation can produce macroscopic phase transitions. Greiner \textit{et al.} have studied BE condensate in the 3D optical lattice where they have seen a phase transition from the superfluid to Mott insulator phase \cite{Greiner}. It is governed by lattice potential where the increase in lattice potential results in the phase transition.
Tosihira Sato \textit{et al.} have shown the phase transition from Abrikosov vortex lattice to pinned vortex lattice \cite{pin}. In case of fast rotating BEC in single planer condensate with dipole-dipole interaction when s-wave interaction becomes attractive and exceeds a critical value then a phase transition occurs which transits triangular lattice to square lattice \cite{sqr}.

In nature sometimes we have seen non-uniform rotation such as rotation of the planets around the Sun, as we move away from the Sun the rotation frequency become less. In the superfluid system, we have little attention on the non-uniform rotation.
Now in this article, we are going to study BEC in a non-uniform rotating system for single species atomic system. We shall describe the system like a superfluid is confined in a container and we are applying rotation in such a way that rotation in center of the condensate is maximum and away from the center the rotation is gradually decreasing.


\section{II. theoretical framework \& Numerical Technique}

The low energy interaction between Bose atoms in momentum space is constant, $U_0 = \frac{4\pi \hbar^2 a}{m}$ with $m$ is the mass of an atom and $a$ the s-wave scattering length (for repulsive interaction $a$ is positive and for attractive interaction it is negative), the Fourier transformation of this interaction in the coordinate space is the delta function potential, contact interaction. In this interaction, the condensate is governed by a nonlinear Schrodinger like equation, known as GP equation, which was first investigated in the superfluid system \cite{pitsing}. It is common practice to solve the GP equation numerically to study the different properties of the condensate \cite{bao, stu GP} in the field of BEC.
Time-dependent GP equation is studied in different geometrical dimensions with isotropic and anisotropic trapping potentials with different interactions between the atoms like spin-orbit interactions, dipole-dipole interactions \cite{sadhan1, sadhan prog, c sadhan, spino, poschl, morse, dipole}. Rotating BEC has been studied taking account of dipolar and spin-orbit interactions \cite{rot dipole, rot spin}.

The GP equation of a condensate, rotating about z-axis with $\Omega$ angular velocity is 
\small
\begin{eqnarray}
  i\hbar\frac{\partial\psi(\textbf{x},t) }{\partial t}=\left(-\frac{\hbar^2}{2m}\nabla^2 + V(\textbf x) + NU_0|\psi|^2 - \Omega L_z\right )\psi(\textbf{x},t)
\end{eqnarray}
\normalsize
where $N$ is the number of atoms in the condensate and $L_z = xp_y - yp_x = -i\hbar (x\partial _y - y\partial _x)$  is the $z$ component of angular momentum. We have considered harmonic trapping $V(\textbf{x}) = \frac{1}{2} m({\omega}_x^2x^2+{\omega}_y^2y^2)$ where $\omega_x, \omega_y$ being the trap frequencies in the $x, y$ directions, and trapping frequency along the z-axis is very high such that the condensate will confine in the xy-plane. Also we have to normalize the wave-function by
\begin{equation}
  \int_{R_d} |\psi(\textbf{x},t)|^2 d\textbf{x} = N
\end{equation}
 We consider a cylindrical symmetric condensate for which we have $\omega_x = \omega_y$.  We transform the variables of the GP equation in dimensionless parameters, to make it numerically convenient.  We take the transformation of variables as $t \rightarrow {\omega}t, \textbf{x} \rightarrow \frac {\textbf{x}} {x_s}$, $\Omega \rightarrow \Omega/\omega, \psi(\textbf{x},t) \rightarrow {x^{3/2}_s} \psi(\textbf{x},t) $\cite{bao, wbao rot} where $x_s$ is the characteristic length of the condensate. Plugging these values in equation (1) and  multiplying by $1/m\omega^2{x^2_s}$ we get
\small
\begin{eqnarray}
  i\varepsilon \frac{\partial\psi(\textbf{x},t) }{\partial t}=\left(-\frac{\varepsilon^2}{2}\nabla^2 + V(\textbf x) + \delta {\varepsilon}^{5/2}|\psi|^2 - \Omega L_z\right )\psi(\textbf{x},t)
\end{eqnarray}
\normalsize
\begin{equation}
\varepsilon = \frac {\hbar}{m{\omega}x_s^2} = \left(\frac {a_0} {x_s}\right)^2
\end{equation}
\begin{equation}
\delta = \frac {NU_0} {a_0^3\hbar\omega} ,    a_0 = \sqrt \frac{\hbar} {m\omega}
\end{equation}
The coefficient of the nonlinearity i.e. the interaction strength parameter
\begin{equation}
g = \delta {\varepsilon}^{5/2} = {\frac {4\pi aN}{a_0}} {\left(\frac {a_0} {x_s}\right)^5}
\end{equation}
So the GP equation takes the form in the dimensionless variables
\begin{equation}
   i\frac{\partial\psi }{\partial t}=\left(-\nabla ^2 + \frac{r^2}{2} + g|\psi|^2 - \Omega L_z\right )\psi
\end{equation}

A typical set of parameters used in experiments with $^8$$^7$Rb is given by $m = 1.44 \times 10^{-25}$ (kg), $\omega = 20\pi$ (rad/sec), $a = 5.1 \times 10^{-9}$ (m) and $\hbar = 1.05 \times 10^{-34}$ (J.s). If one chooses $x_s = a_0$ then $g = \delta$ and the relation between $g$ and $N$ using those parameters $N \approx 53.155g$.

In our case, the rotation is maximum at center and it starts to decrease in radially outward directions. In case of uniformly rotating condensate the angular momentum operator ($L_z$) is multiplied only by angular velocity ($\Omega$) whether in our case there is an additional term $\lambda e^{-{\frac{r^2}{2}}}$ along with angular velocity term ($\Omega$). Now to describe such kind of system the corresponding GP equation will take it the dimensionless form like
\begin{equation}
   i\frac{\partial\psi }{\partial t}=\left[-\nabla ^2 + \frac{r^2}{2} + g|\psi|^2 - (\Omega + \lambda e^{-{\frac{r^2}{2}}}) L_z\right ]\psi
\end{equation}
Where $\lambda e^{-{\frac{r^2}{2}}}$ describes the non-uniform rotating term which is decreasing radially from the center of the condensate.

\begin{figure}[htbb]
  \begin{center}
      \includegraphics[width=3.7cm]{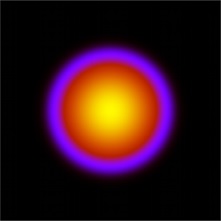} \includegraphics[width=3.7cm]{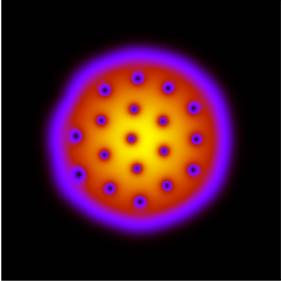}
    \includegraphics[width=2.5cm]{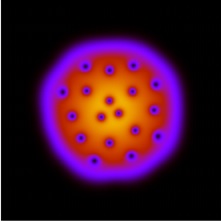}
 \includegraphics[width=2.5cm]{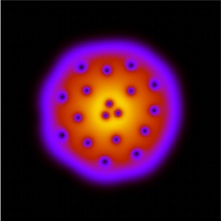}
\includegraphics[width=2.5cm]{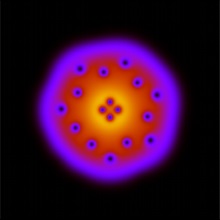}
      \includegraphics[width=2.5cm]{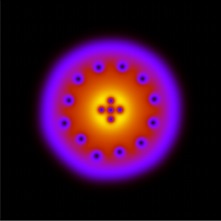}
 \includegraphics[width=2.5cm]{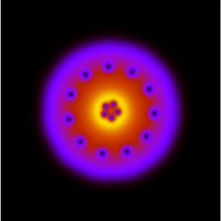}
 \includegraphics[width=2.5cm]{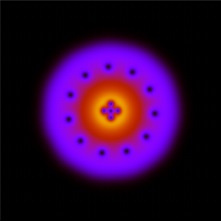}
\caption{(color online) Surface plots of ground state density of the condensate $|\psi(x,y)|^2$ for the system with interaction parameter $g = 1000$. Gaussian type of distribution is observed for non-rotating (top-left) and triangular vortex lattice is observed for the uniform rotation, $\lambda=0$ (top-right),  from second row in text sequence, the density profile of the condensate has been shown for different values of $\lambda$, starting from 1 to 6 with an increment of 1 with $\Omega = 0.70$ as fixed value. It is clear that there is a sharp circular ring shape pattern at the boundary at around $\lambda=5$.}
    \label{fig:1}
  \end{center}
\end{figure}

The system is in the stationary state with a constant rotation about the z-axis. We have solved the time dependent GP equation by backward Euler method to get the stationary state with some suitable interaction parameter $g = 500,\; 1000,\; 2000$. We start with a Gaussian function $\psi(x,y) = exp[{-{\frac{(x^2+y^2)}{2}}}]$ as initial guess and find the solution after one million iteration with time step 0.001 sec for a square geometry of are $[20 \times 20]$ in natural unit
After getting the state $\psi(x,y)$ we have calculated the energy and chemical potential ($\mu$) of the system.
 
\begin{eqnarray}
E = \int _{R_d}  ( \frac{1}{2}|\nabla \psi| ^2 &+& V(x)\psi(x) + \frac{g}{2}|\psi|^4 \\
&-&(\Omega+\lambda e^{-{\frac{r^2}{2}}}) \Re (\psi^*L_z\psi)  ) d\textbf{x} \nonumber
\end{eqnarray}
here $\Re$ represents the real part of a function.
\begin{equation}
\mu = E + \frac{g}{2} \int _{R_d}  |\psi|^4 d\textbf{x}
\end{equation}

\section{III. Results \& discussions }

In FIG. 1, FIG. 2 and FIG. 3 we have plotted the density profile for a range of $\lambda$ values at a fixed value of angular velocity $\Omega$ for different interaction parameters $g=1000$, 2000 and 500 respectively for a system size of 20$\times$20.

\begin{figure}[htbb]
  \begin{center}
    \includegraphics[width=2.5cm]{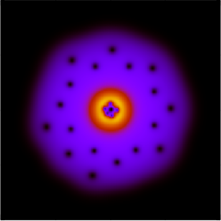}
 \includegraphics[width=2.5cm]{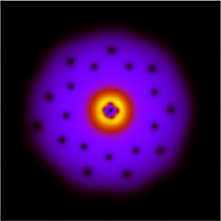}
\includegraphics[width=2.5cm]{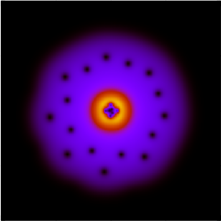}
      \includegraphics[width=2.5cm]{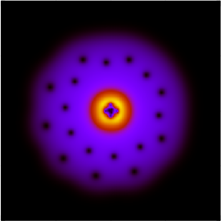}
 \includegraphics[width=2.5cm]{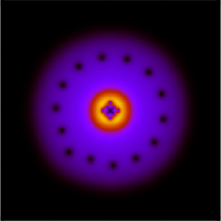}
 \includegraphics[width=2.5cm]{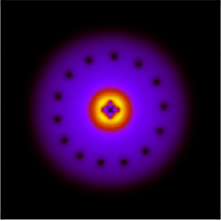}
\caption{(colour online) Surface plots of ground state density of the condensate $|\psi(x,y)|^2$ for the system with interaction parameter $g =2000$ for different $\lambda$ values. The density profile of the condensate has been shown in text sequence for different values of $\lambda$, starting from 12 to 17 with an increment of 1 with $\Omega = 0.70$ as fixed value. It is clear that there is a sharp circular ring shape pattern at the boundary at around $\lambda=17$. }
    \label{fig:2}
  \end{center}
\end{figure}

\begin{figure}[htbb]
  \begin{center}
    \includegraphics[width=2.5cm]{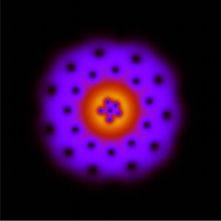}
 \includegraphics[width=2.5cm]{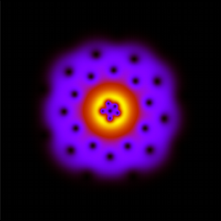}
\includegraphics[width=2.5cm]{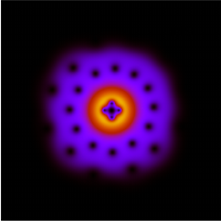}
      \includegraphics[width=2.5cm]{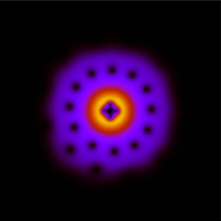}
 \includegraphics[width=2.5cm]{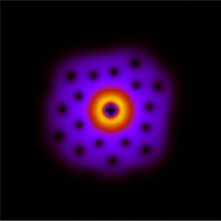}
 \includegraphics[width=2.5cm]{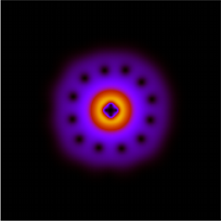}
\caption{(colour online) Surface plots of ground state density of the condensate $|\psi(x,y)|^2$ for the system with interaction parameter $g =500$ for different $\lambda$ values. The density profile of the condensate has been shown in text sequence for different values of $\lambda$, starting from 5 to 10 with an increment of 1 with $\Omega = 0.90$ as fixed value. It is clear that there is a sharp circular ring shape pattern at the boundary at around $\lambda=10$. }
    \label{fig:2}
  \end{center}
\end{figure}

The triangular symmetry of the vortex lattice destroys in the presence of non-uniform rotation, irrespective of the value of the non-uniform rotation parameter $\lambda$. Now if we increase the value of $\lambda$, we see that at a particular value of $\lambda$ we represent this value as $\lambda_c$ the vortices arrange themselves in a ring shape pattern at the boundary of the condensate and remain there if we further increase the value of $\lambda$ at a fixed value of $\Omega$. So we have two phases, one is the disordered phase below the critical non-uniform rotation parameter $\lambda_c$ and ring shape arrangement above  $\lambda_c$.

\begin{figure}[htbb]
  \begin{center}
    \includegraphics[width=2.0cm]{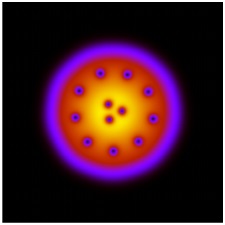} \includegraphics[width=2.0cm]{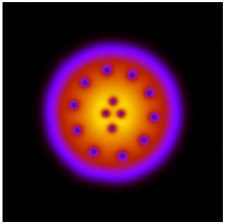} \includegraphics[width=2.0cm]{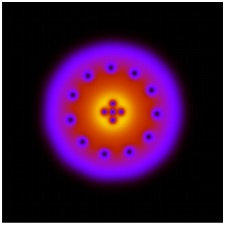} \includegraphics[width=2.0cm]{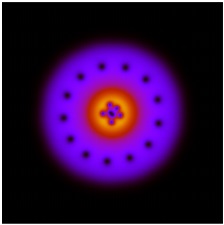}
   
      \includegraphics[width=2.0cm]{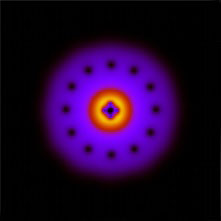} \includegraphics[width=2.0cm]{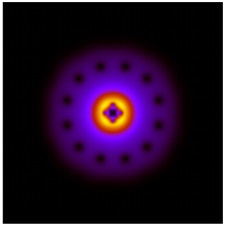} \includegraphics[width=2.0cm]{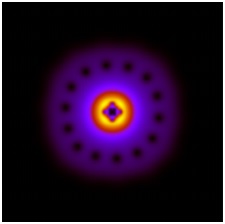} \includegraphics[width=2.0cm]{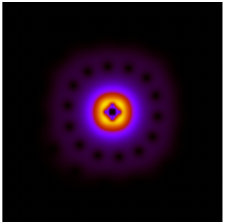}
    \caption{(colour online) Surface plots of ground state density function $|\psi(x,y)|^2$ in 2D $g = 1000$ for different values of $\Omega$ above $\lambda_c$ (starting from $\Omega = 0.60$ to $\Omega = 0.95$ with an increment of $0.05$ in text sequence) }
    \label{fig:3}
  \end{center}
\end{figure}
\begin{figure}[htbb]
\begin{center}
    \includegraphics[width=2.0cm]{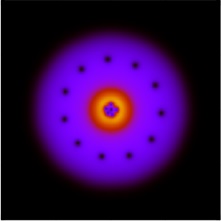} \includegraphics[width=2.0cm]{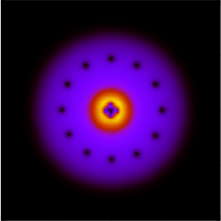} \includegraphics[width=2.0cm]{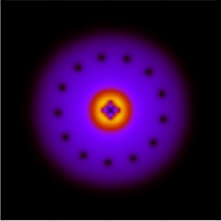} \includegraphics[width=2.0cm]{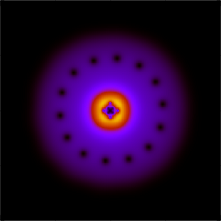}
      \includegraphics[width=2.0cm]{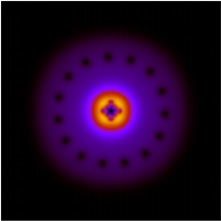} \includegraphics[width=2.0cm]{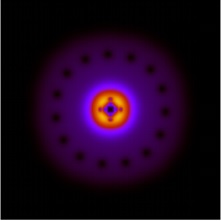} \includegraphics[width=2.0cm]{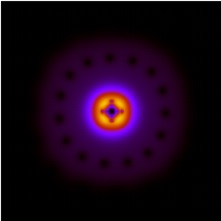} \includegraphics[width=2.0cm]{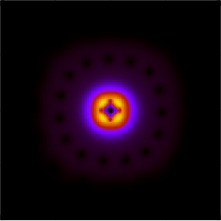}
    \caption{(colour online) Surface plots of ground state density function $|\psi(x,y)|^2$ in 2D $g = 2000$ for different values of $\Omega$ above $\lambda_c$ (starting from $\Omega = 0.60$ to $\Omega = 0.95$ with an increment of $0.05$ in text sequence) }
    \label{fig:4}
  \end{center}
\end{figure}

\begin{figure}[htbb]
\begin{center}
    \includegraphics[width=3.0cm]{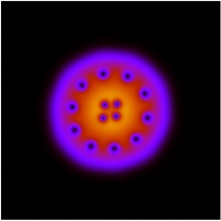} \includegraphics[width=3.0cm]{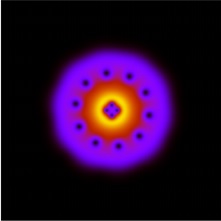} \includegraphics[width=2.5cm]{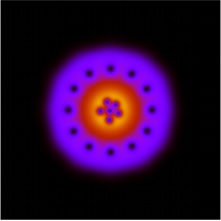}
    \includegraphics[width=2.50cm]{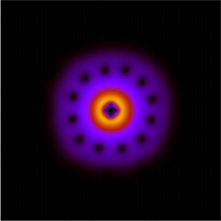} \includegraphics[width=2.5cm]{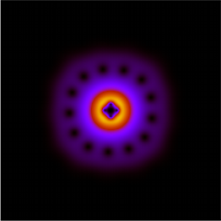}
    \caption{(colour online) Surface plots of ground state density function $|\psi(x,y)|^2$ in 2D $g = 500$ for different values of $\Omega$ above $\lambda_c$ (starting from $\Omega = 0.75$ to $\Omega = 0.95$ with an increment of $0.05$ in text sequence) }
    \label{fig:5}
  \end{center}
\end{figure}

\begin{figure}[htbb]
  \begin{center}
    \includegraphics[width=4.0cm]{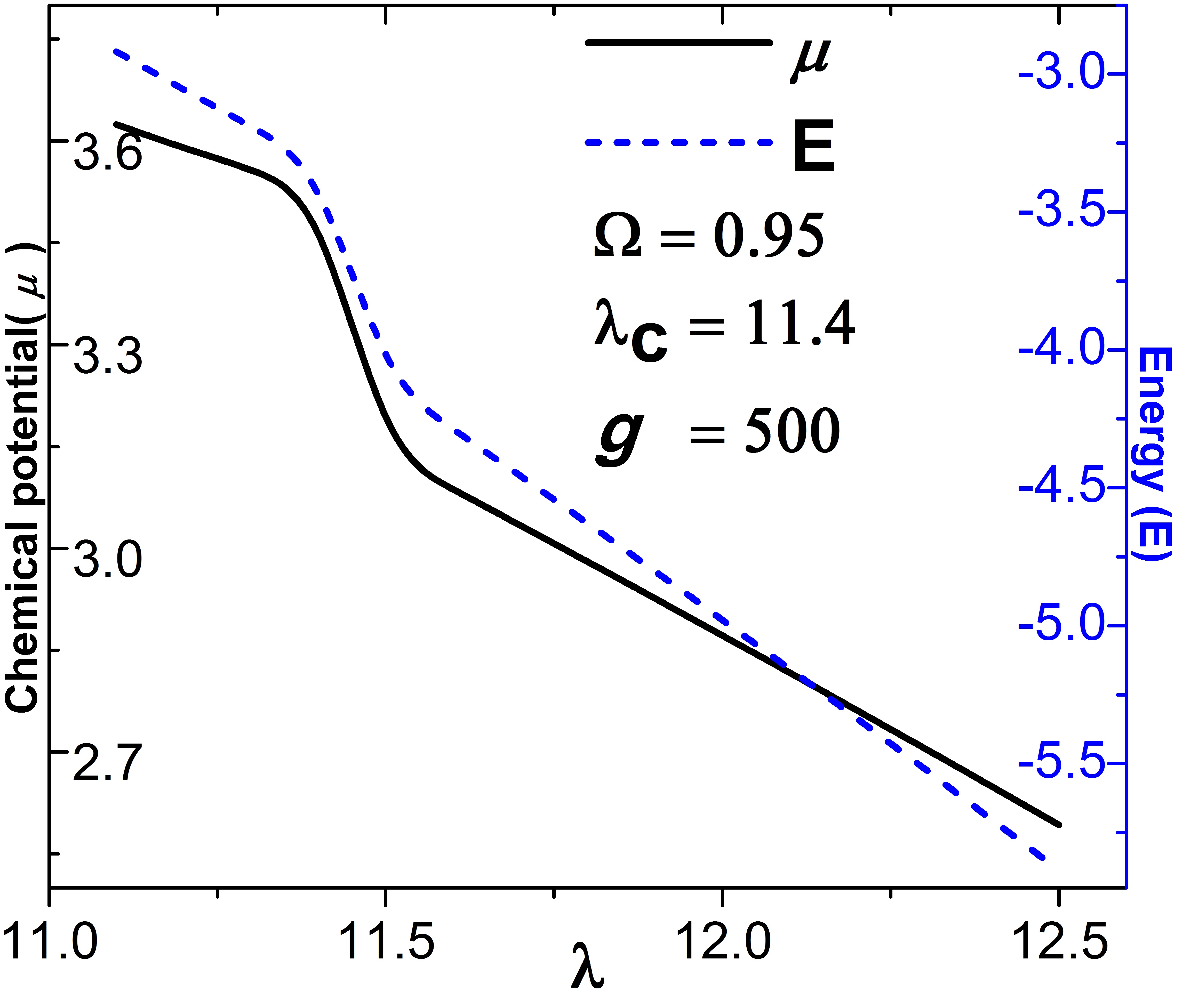}  
      \includegraphics[width=4.0cm]{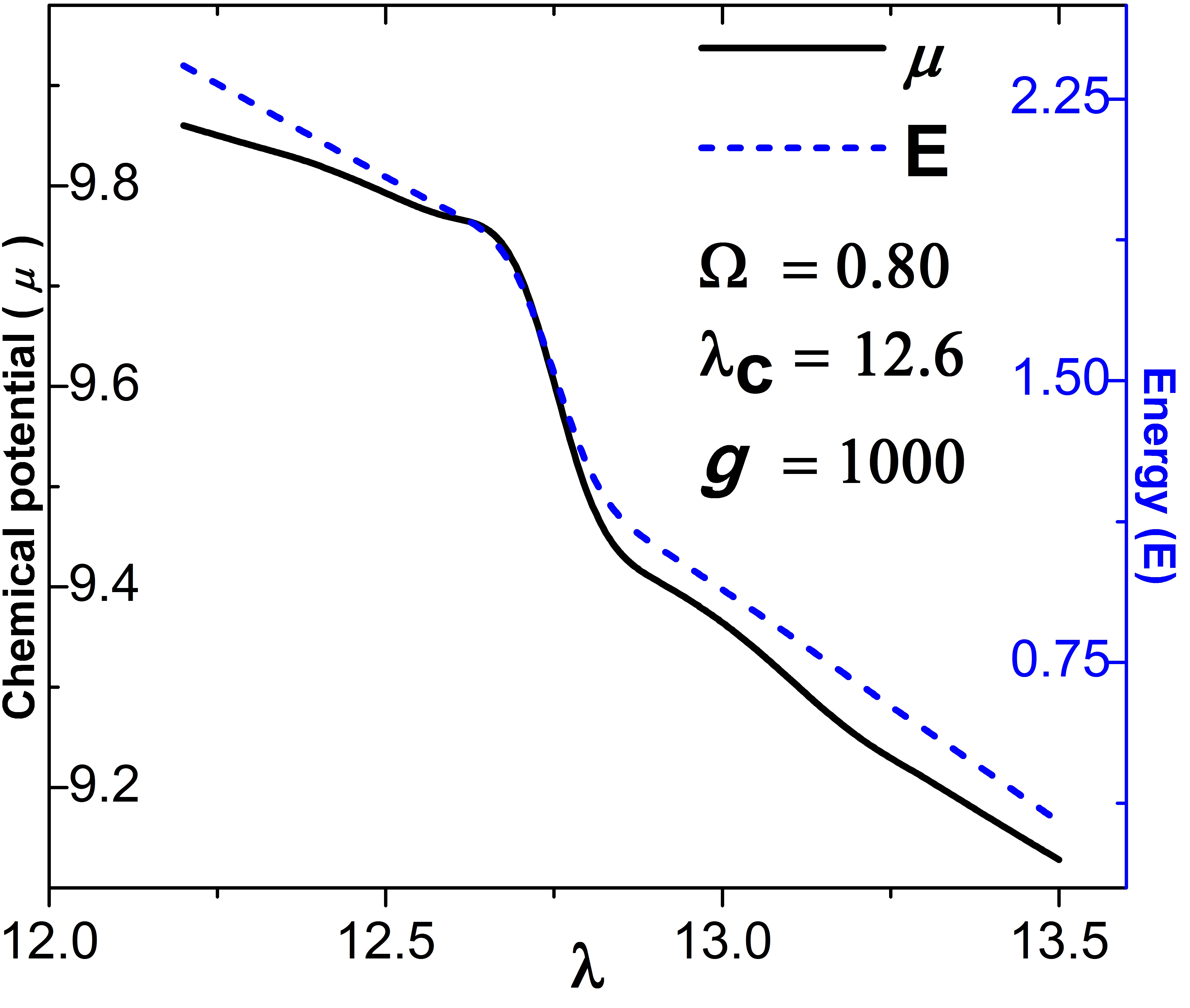}        \includegraphics[width=5.0cm]{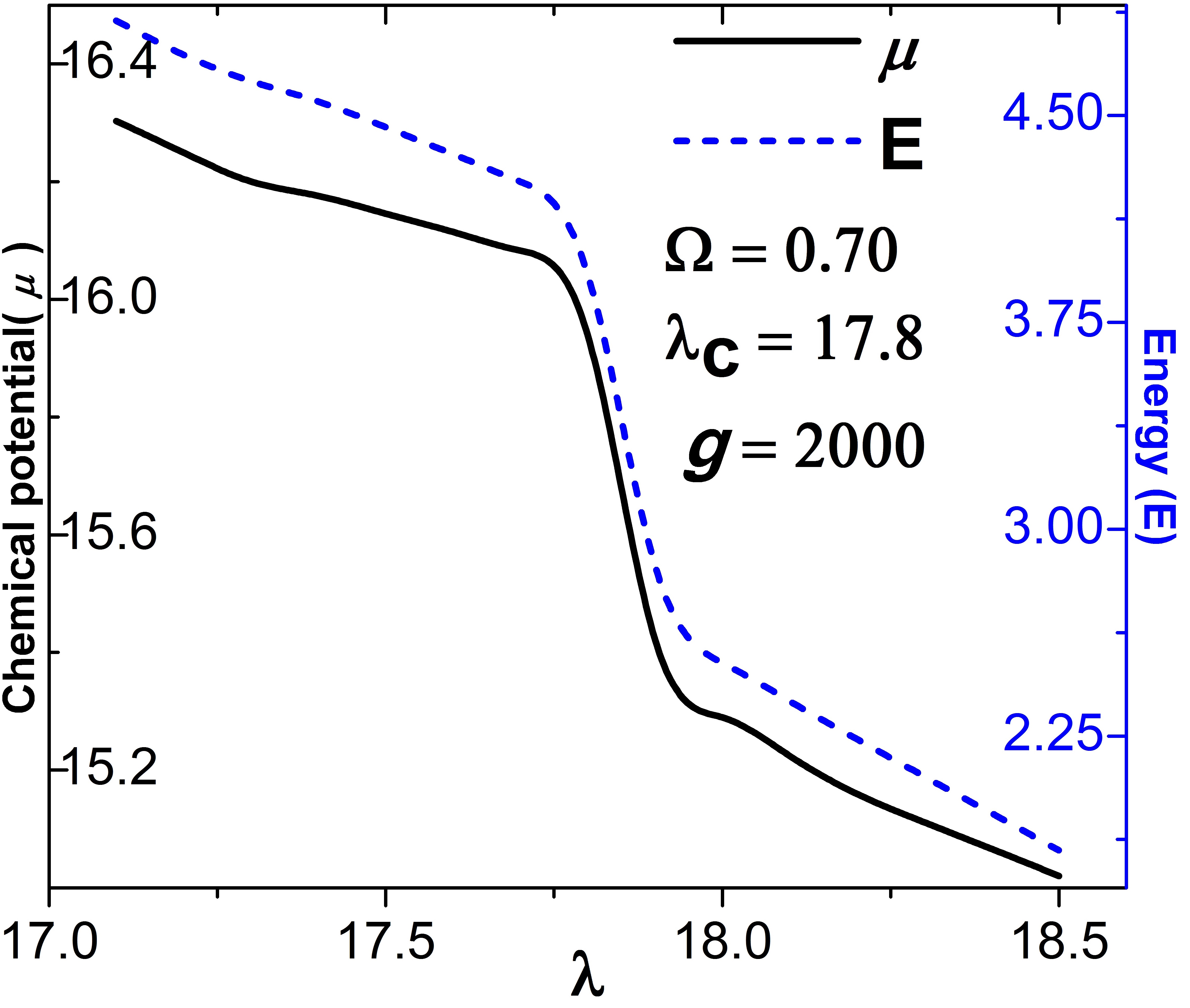}
    \caption{(colour online) Variation of chemical potential ($\mu$) and energy, with $\lambda$ for [$\Omega = 0.95$ $g$ = 500], [$\Omega = 0.80$ $g$ = 1000] and [$\Omega = 0.70$ $g$ = 2000]}
    \label{energy-mu}
  \end{center}
\end{figure}

\begin{figure}
\includegraphics[width=7.2cm]{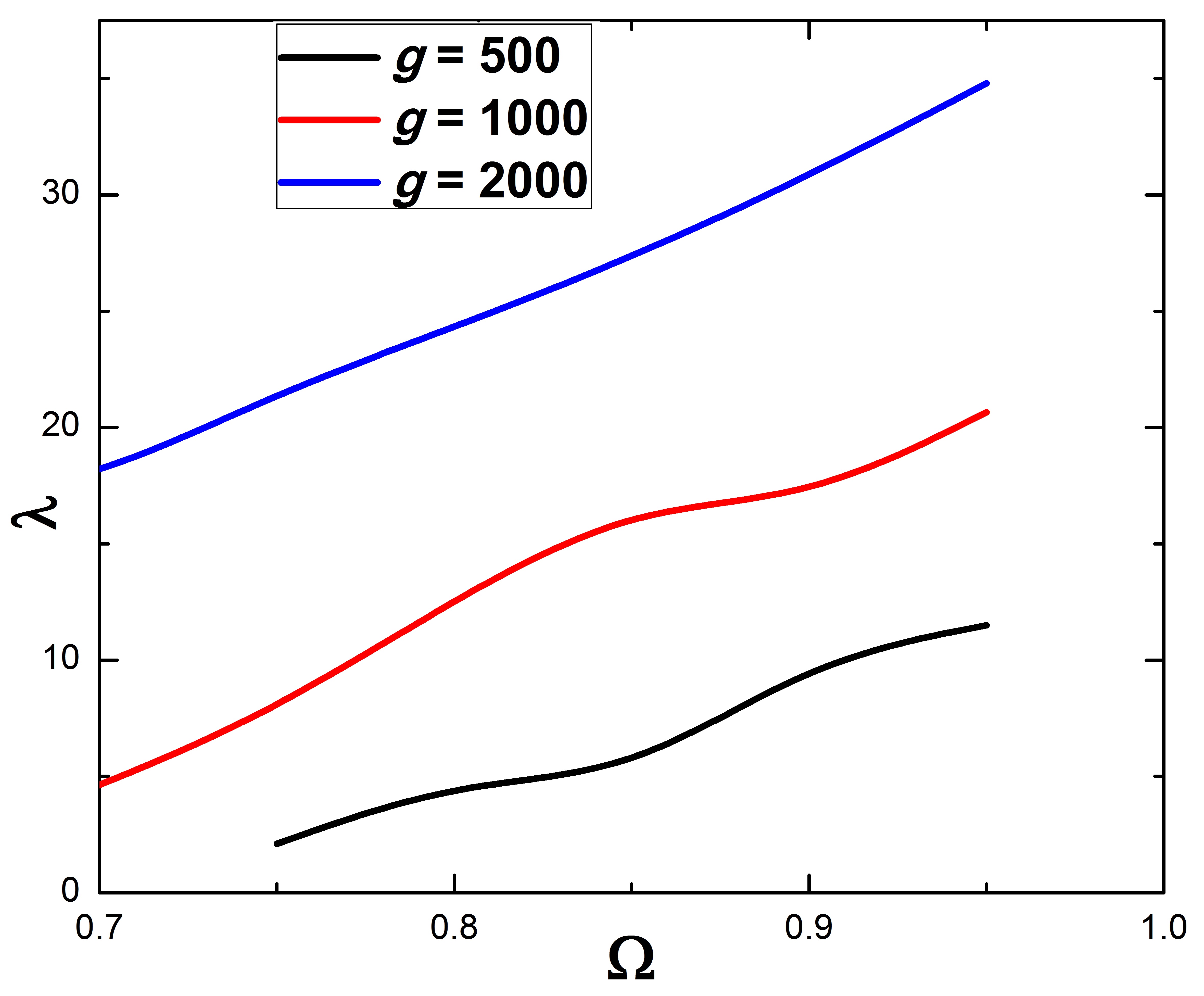}
\caption{(colour online) {\bf Phase diagram:} Below the line we observed the disordered lattice and above the line circular ring shape pattern}
\label{phase}
\end{figure}

In FIG. 4, FIG. 5 and FIG. 6 we have plotted the density profile of the condensate for different values of $\Omega$ with a suitable value of $\lambda$ (above $\lambda_c$). We see that the ring shape pattern remain the same if we increase the value of $\Omega$. These studies show that there is a phase diagram associated with the transition of vortex lattice form disorder to ring shape pattern. We have drawn this diagram in FIG. \ref{phase}, in the high rotation limit and high non-rotating parameter region is the ring shape pattern region and below the line is the disordered region.

To complete the study we have calculated the energy and chemical potential as a function of the non-rotating parameter. We have seen that at phase transition point there is a change of these physical quantities. This is due to the different mass distribution of the condensate. All this study suggest that the presence of non-uniform rotation has a rich phase of vortex lattice. At present, we don't have any experimental result regarding this non-uniform rotation, but there is the possibility to get this in near future.

\end{document}